\documentclass[aps,showpacs, prl,twocolumn,superscriptaddress]{revtex4}%
\usepackage{epsfig,dsfont,amssymb,amsmath,amsthm,amsfonts,amsbsy,mathrsfs}
\usepackage{graphicx}
\usepackage{amsmath}
\usepackage{amssymb}
\begin{document}
\title{Four superhard carbon allotropes: First-principle study}
\author{Chaoyu He}
\affiliation{Institute for Quantum Engineering and Micro-Nano Energy
Technology, Xiangtan University, Xiangtan 411105, China}
\author{L. Z. Sun}
\email{lzsun@xtu.edu.cn} \affiliation{Institute for Quantum
Engineering and Micro-Nano Energy Technology, Xiangtan University,
Xiangtan 411105, China}
\author{C. X. Zhang}
\affiliation{Institute for Quantum Engineering and Micro-Nano Energy
Technology, Xiangtan University, Xiangtan 411105, China}
\author{K. W. Zhang}
\affiliation{Institute for Quantum Engineering and Micro-Nano Energy
Technology, Xiangtan University, Xiangtan 411105, China}
\author{Xiangyang Peng}
\affiliation{Institute for Quantum Engineering and Micro-Nano Energy
Technology, Xiangtan University, Xiangtan 411105, China}
\author{Jianxin Zhong}
\email{zhong.xtu@gmail.com}\affiliation{Institute for Quantum
Engineering and Micro-Nano Energy Technology, Xiangtan University,
Xiangtan 411105, China}
\date{\today}

\begin{abstract}
Using a generalized genetic algorithm, we propose four new sp$^3$
carbon allotropes with 5-6-7 (5-6-7-type Z-ACA and Z-CACB) or 4-6-8
(4-6-8-type Z4-A$_3$B$_1$ and A4-A$_2$B$_2$) carbon rings. Their
stability, mechanical and electronic properties are systematically
studied using first-principles method. We find that the four new
carbon allotropes show amazing stability in comparison with the
carbon phases proposed recently. Both 5-6-7-type ZACA and Z-CACB are
direct-band-gap semiconductors with band gaps of 2.261 eV and 4.196
eV, respectively. However, the 4-6-8-type Z4-A$_3$B$_1$ and
A4-A$_2$B$_2$ are indirect-band-gap semiconductors with band gaps of
3.105 eV and 3.271 eV, respectively. Their mechanical properties reveal that all the four carbon
allotropes proposed in present work are superhard materials comparable to diamond.\\
\end{abstract}
\maketitle
\indent Carbon is considered as the most active element in the periodic table due to its broad sp, sp$^{2}$ and sp$^{3}$ hybridizing ability. Besides the four best-known carbon allotropes, graphite, cubic diamond (C-diamond), hexagonal diamond (H-diamond) and amorphous carbon, an unknown superhard phase of carbon has been reported in experiment \cite{1, 2, 3, 4} along with the structural phase transition in cold compressing graphite. Several structures have been proposed theoretically as the candidate for this superhard phase, such as the monoclinic M-carbon\cite{5}, cubic body center C4 carbon (bct-C4) \cite{6} and the orthorhombic W-carbon \cite{7}. Although the monoclinic one, namely M-carbon, has been preliminarily identified by a following experimental process of compressing graphite\cite{arl}, bct-C4 and orthorhombic W-carbon can also fit the experimental XRD-data to some extent. These discoveries attract great interest in theoretical predication \cite{8, 9, 10} and experimental research \cite{9} on such superhard carbon allotropes. Very recently, another new carbon allotrope, named as Z-carbon, was proposed and investigated at almost the same time by three independent research groups \cite{8, 9, 10} (this structure was also named as oC16II in reference [8] and Cco-C8 in reference [10]). Z-carbon is more stable (its cohesive energy is about 129 meV per atom above diamond) and harder than bct-C4, M-carbon and W-carbon. Moreover, its transition pressure is around 10 Gpa which is lower than those of bct-C4, M-carbon and W-carbon. Thus, it is believed that cold compressing graphite prefers forming Z-carbon. Although none of them can solely fit the experimental results satisfactorily, these theoretically proposed intermediate phases are significant in understanding the experimental process of cold compressing graphite and the cold compressing carbon nanotubes \cite{cnt}. Moreover, these theoretical studies also arouse great interest on the superhard carbon materials \cite{11, 12, 13} and analogical superhard BN phases \cite{14, 15}.\\
\begin{figure}
\centering
\includegraphics[width=3.00in]{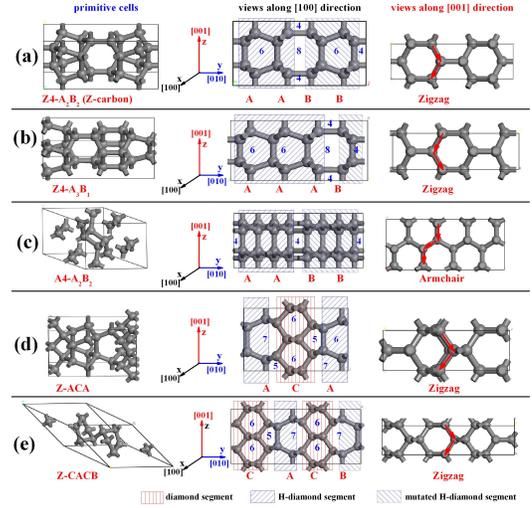}\\
\caption{Primitive cell and corresponding side and top views of
Z-carbon (a), Z4-A$_3$B$_1$ (b), A4-A$_2$B$_2$ (c), Z-ACA (d) and
Z-CACB (e). The numbers in the side views panel are the types of
carbon rings. Shadows indicate how they hybridized from C-diamond ,
H-diamond and mutated H-diamond gene segments.}\label{fig1}
\end{figure}
\indent All above new carbon phases can be designed by using the recently developed particle-swarm optimization method \cite{16}, graph theoretical methods \cite{17} and the evolutionary algorithm USPEX developed by Oganov \cite{uspex}. All of them can be divided into two groups: 5-6-7-type (M-carbon \cite{5} and W-carbon \cite{7} containing 5-, 6-, 7-carbon rings) and 4-6-8-type (bct-C4 \cite{6} and Z-carbon \cite{9} containing 4-, 6-, 8-carbon rings). We notice that, from the point of view of structure, they all can be constructed through mutating H-diamonds or combing the segments of H-diamond and C-diamond. H-diamond and C-diamond are the most favorable sp$^3$ carbon allotropes in nature that could be used as excellent parents for finding new carbon allotropes through hybridizing their stable segments. For example, bct-C4 and Z-carbon can be looked upon as the hybridization of H-diamond and mutated H-diamond. M-carbon and W-carbon can be taken as the hybridization of distorted H-diamond and C-diamond segments. The 4H \cite{18}, 6H \cite{18} and 12R \cite{19} carbon allotropes can be regarded as the superlattice of C-diamond (along [111] direction) and H-diamond (along 001 direction). By hybridizing C-diamond and H-diamond in different manners, almost all previously proposed carbon structures can be obtained. Such a structural construction process is compatible with the essence of genetic algorithm (GA). The genetic algorithm is widely used in searching for zero-dimensional (0D) element clusters such as the carbon fullerenes \cite{GA}. Moreover, it is an effective method in prediction three-dimensional (3D) superhard carbon phases \cite{uspex}. In this paper, using a generalized GA, we choose H-diamond and C-diamond as parents to hybridize new carbon allotropes. Four new carbon allotropes with 5-6-7 (Z-ACA and Z-CACB) or 4-6-8 (Z4-A$_3$B$_1$ and A4-A$_2$B$_2$) carbon rings are proposed in our present work and their stability, electronic and mechanical properties are systematically studied using first-principles calculations based on the density functional theory. All the four carbon allotropes in our present work are more favorable than bct-C4. Our results indicate that all of them are superhard insulators with direct or indirect band gaps. Because the four new allotropes are more stable than graphite under certain pressures, all of them are expected to be obtained from cold compressing graphite.\\
\begin{table*}
  \centering
  \caption{GGA calculated structure information for Z-carbon, Z4-A$_3$B$_1$, A4-A$_2$B$_2$, Z-ACA and Z-CACB at 0 GPa.}\label{tabI}
\begin{tabular}{c c c c}
\hline
System &Space group and cell &inequivalent atoms position &rings-type\\
\hline
Z-carbon &Cmmm (NO*65)                                 &(0.089, 0.316, 0.500)   &4-6-8 \\
         &a=8.772 {\AA}, b=4.256 {\AA}, c=2.514 {\AA}  &(0.167, 0.185, 1.000)   &      \\
Z4-A$_3$B$_1$   &Pmmn (NO*59)                          &(0.041, 0.312, 0.500)   &4-6-8 \\
         &a=8.762 {\AA}, b=4.263 {\AA}, c=2.514 {\AA}  &(0.208, 0.185, 0.500)   &      \\
         &                                             &(0.285, 0.316, 1.000)   &      \\
         &                                             &(0.464, 0.318, 1.000)   &      \\
A4-A$_2$B$_2$   &Cmca (NO*64)                          &(0.317, 0.067, 0.088)   &4-6-8 \\
         &a=4.257 {\AA}, b=10.114 {\AA} c=4.363 {\AA}  &(0.314, 0.188, 0.584)   &      \\
Z-ACA     &Pmmn (NO*59)                                &(0.000, 0.500, -0.069)  &5-6-7 \\
         &a=4.760 {\AA}, b=2.521 {\AA}, c=7.930 {\AA}  &(-0.500, 0.500, -0.043) &      \\
         &                                             &(-0.265, 0.500, -0.171) &      \\
         &                                             &(-0.326, 1.000, -0.276) &      \\
         &                                             &(-0.196, 1.000, -0.453) &      \\
Z-CACB  &Imma (NO*74)                                  &(0.500, 0.750, 0.542)   &5-6-7 \\
         &a=4.876 {\AA}, b=2.529 {\AA}, c=11.535 {\AA} &(0.237, 0.750, 0.618)   &      \\
         &                                             &(0.177, 0.250, 0.691)   &      \\
         &                                             &(0.500, 0.750, 0.965)   &      \\
\hline
\end{tabular}
\end{table*}
\section{Models and Methods}
\subsection{Models}
\indent To compare with previously proposed carbon allotropes, we restrict the atoms in the primitive unit cell no larger than 16 carbon atoms. The primitive cells and the side and top views of Z-carbon, Z4-A$_3$B$_1$, A4-A$_2$B$_2$, Z-ACA and Z-CACB are shown in Fig.~\ref{fig1}. Here Z and A denote that the framework of the systems are constructed with zigzag and armchair carbon chains along [100] direction, respectively. From the side view, we can see that Z-carbon (Z4-A$_2$B$_2$, as shown in Fig.~\ref{fig1} (a)) is composed of two A gene segments derived from H-diamond and two B gene segments derived from mutated H-diamond (the mutation is a c/2 translation along the [001] orientation). At the AA-BB interface, a series of 4 and 8 carbon rings appear. According to such denomination, the previously proposed bct-C4 is Z2-A$_1$B$_1$ (Z4-A$_1$B$_1$A$_1$B$_1$) due to the fact that there are one A gene segment and one B gene segment in its crystal cell. Z4-A$_3$B$_1$ as shown in Fig.~\ref{fig1}(b) is hybridized by a triple A gene segment derived from H-diamond and one B segment derived from mutated H-diamond. The primitive cell of A4-A$_2$B$_2$ contains 16 carbon atoms and its crystal cell contains one double A gene segment and one double B segment of H-diamond, as shown in Fig.~\ref{fig1}(c). Z3- systems are restricted according to the topological requirement. The systems of A3-, A4-A$_3$B$_1$, Zn- and An- (n is larger than 4) containing more than 16 atoms in the primitive cell are not included in present work. The systems of A2-A$_1$B$_1$ and A4-A$_1$B$_1$A$_1$B$_1$ are energetically unstable in comparison with bct-C4. In Fig.~\ref{fig1} (d) and (e), we show the situations of hybridizing H-diamond and C-diamond in the zigzag direction with 5-7 carbon rings at the C/A or C/B interface. The system by hybridizing H-diamond and C-diamond in the armchair direction is unstable because it is contained non-four-connected carbon atoms. In fact, by hybridizing H-diamond and C-diamond segments, many new allotropes can also be constructed. In present work, two low energy allotropes named as Z-ACA and Z-CACB containing 16 and 12 atoms in their primitive cell, respectively, are taken as examples. Here, A, B and C denote the H-diamond, mutated H-diamond and C-diamond gene segment, respectively.\\
\subsection{Methods}
\indent To investigate the stability, electronic and mechanical properties of these new carbon allotropes, first-principles calculations based on the density functional theory are employed. All calculations are performed within general gradient approximation (GGA) \cite{20} as implemented in Vienna ab initio simulation package (VASP) \cite{21, 22}. The interactions between nucleus and the 2s$^{2}$2p$^{2}$ valence electrons of carbon are described by the projector augmented wave (PAW) method \cite{23, 24}. A plane-wave basis with a cutoff energy of 500 eV is used to expand the wave functions. The Brillouin Zone (BZ) sample meshes are set to be dense enough (less than 0.21/{\AA}) to ensure the accuracy of our calculations. Crystal lattices and atom positions of graphite, diamond, bct-C4, M-carbon, W-carbon, Z-carbon, Z4-A$_3$B$_1$, A4-A$_2$B$_2$, Z-ACA and Z-CACB are fully optimized up to the residual force on every atom less than 0.005 eV/{\AA} through the conjugate-gradient algorithm. Vibration properties are calculated by using the phonon package \cite{25} with the forces calculated from VASP. To evaluate the transition pressure from graphite to superhard phase, the exchange-correlation functional is describe by LDA \cite{26, 27} for the consideration that LDA can give reasonable interlayer distances, mechanical properties of graphite sheets due to a delicate error cancelation of the exchange and correlation interactions in comparison with that of semi-local generalized gradient approximation (GGA).\\
\begin{figure}
\includegraphics[width=3.0in]{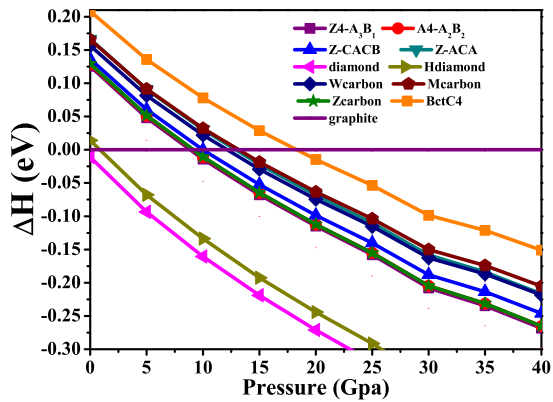}\\
\caption{Enthalpy per atom for C-iamond, H-diamond, bct-C4,
M-carbon, W-carbon, Z-carbon, Z4-A$_3$B$_1$-carbon, A4-A$_2$B$_2$,
Z-ACA and Z-CACB as a function of pressure relative to graphite
derived from LDA calculations.}\label{fig2}
\end{figure}
\section{Results and Discussion}
\subsection{Structures}
\indent Similar to bct-C4 and Z-carbon, Z4-A$_3$B$_1$ and A4-A$_2$B$_2$ belong to 4-6-8-type. The crystal structure of Z4-A$_3$B$_1$ belongs to Pmmn space group. At zero pressure, its equilibrium lattice constants calculated from GGA are a=8.762 {\AA}, b=4.263 {\AA} and c=2.514 {\AA}. Four inequivalent atoms (16 atoms per primitive cell) in its unit cell occupy the positions at (0.041, 0.312, 0.5), (0.208, 0.185, 0.5), (0.285, 0.316, 1.0) and (0.464, 0.318, 1.0), respectively. A4-A$_2$B$_2$ belongs to Cmca space group and its equilibrium lattice constants obtained from GGA are a=4.257 {\AA}, b=10.114 {\AA} and c=4.363 {\AA}. There are only two inequivalent atoms (16 atoms per primitive cell) in A4-A$_2$B$_2$ locating at positions of (0.317, 0.067, 0.088) and (0.314, 0.188, 0.584). Recently proposed Z-carbon holds Cmmm symmetry with equilibrium lattice parameters a=8.772 {\AA}, b=4.256 {\AA} and c=2.514 {\AA}. There are only two inequivalent atoms in Z-carbon unit cell locating at (0.089, 0.316, 0.5), and (0.167, 0.185, 1.0). Z-ACA and Z-CACB belong to 5-6-7-type similar to M-carbon and W-carbon containing 5-, 6-, and 7-carbon rings. Z-ACA belongs to Pmmn space group and contains five inequivalent atoms (16 atoms per primitive cell) in its orthorhombic lattice with constants of a=4.760 {\AA}, b=2.521 {\AA} and c=7.93 {\AA}. The five inequivalent atom positions are listed in Tab.~\ref{tabI}. Z-CACB with an orthorhombic lattice (a=4.876 {\AA}, b=2.529 {\AA}, c=11.535 {\AA}) belongs to Imma space group containing only 12 carbon atoms per primitive cell with four inequivalent atoms locating at (0.500, 0.750, 0.542), (0.237, 0.750, 0.618), (0.177, 0.250, 0.691), (0.500, 0.750, 0.965). The structure information of Z-carbon, Z4-A$_3$B$_1$, A4-A$_2$B$_2$, Z-ACA and Z-CACB derived from GGA calculations is listed in Tab.~\ref{tabI}. All these new carbon allotropes can be constructed by hybridizing C-diamond and H-diamond as sketched in Fig.~\ref{fig1}.(a)-(e). Moreover, similar to the previously proposed bct-C4, Z-carbon, M-carbon and W-carbon, all of them can be considered as potential products in the process of cold compressing graphite.\\
\begin{figure}
\includegraphics[width=3.0in]{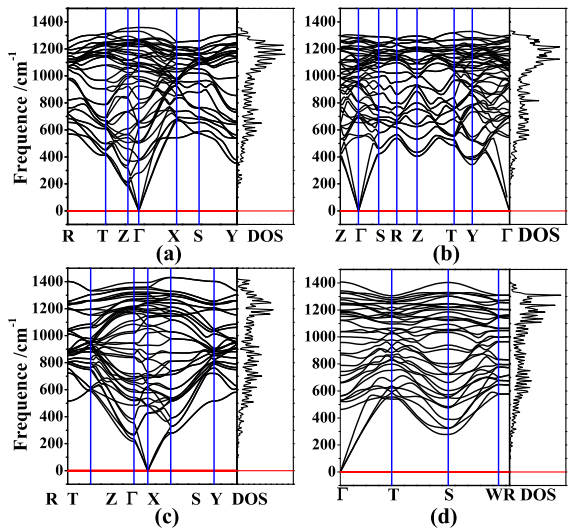}
\caption{Phonon band structure and phonon density of state for
Z4-A$_3$B$_1$ (a), A4-A$_2$B$_2$ (b), Z-ACA (c) and Z-CACB (d) at
zero pressure based on GGA calculations.}\label{fig3}
\end{figure}
\subsection{Stability}
\indent The relative stability of diamond, bct-C4, M-carbon, W-carbon, Z-carbon, Z4-A$_3$B$_1$, A4-A$_2$B$_2$, Z-ACA and Z-CACB is evaluated through comparing their cohesive energy per atom. All these four new allotropes are more energically stable than bct-C4. At zero pressure, the cohesive energy of Z-ACA is about 30 meV lower than bct-C4 and only 5 meV higher than that of M-carbon. Z-CACB is more favorable than both M-carbon and W-carbon. Its cohesive energy is -7.556 eV per atom that is about 17 meV lower than that of W-carbon (-7.539 eV per atom) and 8 meV larger than that of Z-carbon (7.564 eV per atom). The most stable two allotropes are Z4-A$_3$B$_1$ and A4-A$_2$B$_2$ and their cohesive energies are -7.568 eV and -7.565 eV per atom, respectively. Both of them are more stable than Z-carbon and they are the most stable new carbon phases theoretically predicated so far. The enthalpy per atom for diamond, H-diamond, bct-C4, M-carbon, W-carbon, Z-carbon, Z4-A$_3$B$_1$, A4-A$_2$B$_2$, Z-ACA and Z-CACB as a function of pressure relative to graphite derived from LDA calculations is shown in Fig.~\ref{fig2}. The results indicate that Z-ACA is more stable than M-carbon under external pressure and becomes more stable than W-carbon when the pressure is above 35 Gpa. The transition pressures for M-carbon, W-carbon and Z-ACA are very close to each other (located at around 12.1-13.3 Gpa). Z-CACB is always more favorable than bct-C4, M-carbon, W-carbon and Z-ACA and is more stable than graphite when the external pressure is larger than 10 GPa. Z-carbon, Z4-A$_3$B$_1$ and A4-A$_2$B$_2$ have almost the same relative stability and transition pressure. They are more stable than graphite when the external pressure is larger than 9.16 Gpa. To further confirm the dynamic stability of Z4-A$_3$B$_1$, A4-A$_2$B$_2$, Z-ACA and Z-CACB, their phonon band structures and phonon density of states are calculated. The phonon band structures and phonon density of states at zero pressure are show in Fig.~\ref{fig3}. There is no negative frequency for all of the four new carbon allotropes up to 40 Gpa, confirming that these allotropes are dynamic stable phases of carbon. \\
\begin{table}
  \centering
  \caption{Density(g/cm$^{3}$), band gap (Eg: eV), cohesive energy (Ecoh: eV),
  bulk modulus (B0: Gpa) and Vickers hardness (H: Gpa) for diamond, bct-C4, M-carbon,
  W-carbon, Z-carbon, Z4-A$_3$B$_1$, A4-A$_2$B$_2$, Z-ACA and Z-CACB.}\label{tabII}
\begin{tabular}{c c c c c c c }
\hline
Systems     &Density   &Eg       &Ecoh     &B0    &H\\
\hline
diamond     &3.496  &4.635(I)  &-7.693  &473.72   &88.31\\
bct-C4      &3.309  &2.491(I)  &-7.497  &401.91   &67.59\\
M-carbon    &3.336  &3.493(I)  &-7.531  &404.58   &79.24\\
W-carbon    &3.346  &4.281(I)  &-7.539  &400.29   &79.08\\
Z-carbon    &3.399  &3.273(I)  &-7.564  &415.83   &81.09\\
Z4-A$_3$B$_1$      &3.398  &3.105(I)  &-7.568  &415.49   &80.54\\
A4-A$_2$B$_2$      &3.397  &3.271(I)  &-7.565  &413.62   &83.18\\
Z-ACA        &3.353  &2.261(D)  &-7.526  &375.97   &78.74\\
Z-CACB      &3.365  &4.196(D)  &-7.556  &405.41   &82.01\\
\hline
\end{tabular}
\end{table}
\subsection{Mechanical and electronic properties}
\indent Mass density, band gaps, cohesive energies, bulk modulus and hardness of diamond, bct-C4, M-carbon, W-carbon, Z-carbon, Z4-A$_3$B$_1$, A4-A$_2$B$_2$, Z-ACA and Z-CACB are summarized in Tab.~\ref{tabII}. The results indicate that all above carbon allotropes are superhard intermediate phases between graphite and diamond due to their considerable bulk modulus and hardness. Bulk modulus (B0) is obtained by fitting the total energy as a function of volume to the third-order Birch-Murnaghan equation of state. Further hardness evaluation is considered according to the recently introduced empirical scheme\cite{11} which correlates the Vickers hardness to the bulk modulus (B${_0}$) and shear modulus (G) through the formula: H${_v}$=2(G${^3}$/B${^2_0}$)${^{0.585}}$-3. From Tab.~\ref{tabII} we can see that Z-carbon, Z4-A$_3$B$_1$ and A4-A$_2$B$_2$ hold almost the same stability, band gap, mass density, bulk modulus and hardness. The values of bulk modulus are 415.83 Gpa, 415.49 Gpa and 413.62GP for Z-carbon, Z4-A$_3$B$_1$ and A4-A$_2$B$_2$, respectively. The values of Vickers hardness are 81.09 Gpa, 80.54 Gpa and 83.18 Gpa for Z-carbon, Z4-A$_3$B$_1$ and A4-A$_2$B$_2$, respectively, which are comparable to that of diamond (88.31 Gpa). M-carbon, W-carbon, Z-ACA and Z-CACB hold similar stability, density, bulk modulus and Vickers hardness. Their calculated values of Vickers hardness (79.24 Gpa, 79.08 Gpa, 78.74 Gpa and 82.01 Gpa, respectively) are also comparable to the value for diamond. The results indicate that the four new carbon allotropes proposed in present work are superhard materials comparable to diamond. \\
\begin{figure}
\includegraphics[width=3.0in]{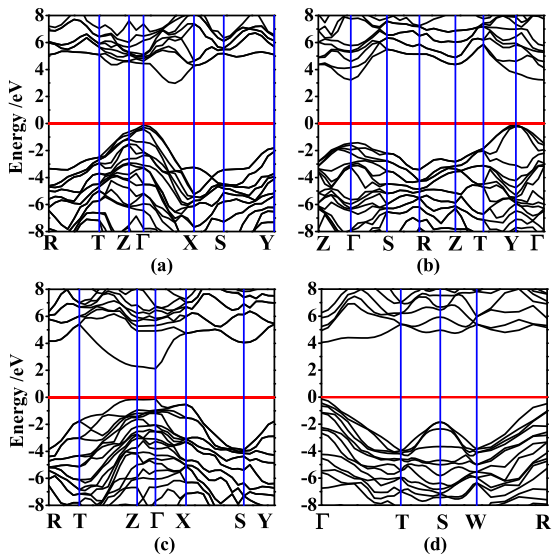}\\
\caption{Electronic band structure of Z4-A$_3$B$_1$ (a),
A4-A$_2$B$_2$ (b), Z-ACA (c) and Z-CACB (d) at zero pressure based
on GGA calculations.}\label{fig4}
\end{figure}
\indent Electronic properties of diamond, bct-C4, M-carbon, W-carbon, Z-carbon, Z4-A$_3$B$_1$, A4-A$_2$B$_2$, Z-ACA, Z-CACB are investigated and the band structures of Z4-A$_3$B$_1$, A4-A$_2$B$_2$, Z-ACA and Z-CACB are shown in Fig.~\ref{fig4}. The results indicate that all these superhard carbon allotropes are indirect wide band gap semiconductors except for Z-ACA and Z-CACB. From the results of Fig.~\ref{fig4} we can see that Z4-A$_3$B$_1$ is an indirect-wide-gap semiconductor and its band gap (3.105 eV ) is larger than that of bct-C4 (2.491 eV) and close to that of Z-carbon (3.273 eV). A4-A$_2$B$_2$ is also an indirect-band-gap semiconductor and its band gap (3.271 eV) is close to those of Z-carbon and Z4-A$_3$B$_1$. Different from diamond, bct-C4, M-carbon, W-carbon, Z-carbon, Z4-A$_3$B$_1$, and A4-A$_2$B$_2$, both Z-ACA and Z-CACB are direct-band-gap semiconductors with gaps of 2.261 eV and 4.196 eV, respectively. The wide band gaps of all these four new carbon allotropes indicate that all of them are transparent carbon phases.\\
\section{Conclusion}
\indent  In summary, using a generalized genetic-algorithm, we proposed four new carbon allotropes. The stability, electronic and mechanical properties of the four new carbon allotropes are investigated using first-principles method. The dynamic stability of all these new carbon phases is confirmed from the phonon band calculations. Under proper external pressure, these four new allotropes of carbon are expected to be obtained from cold compressing graphite. Z4-A$_3$B$_1$ and A4-A$_2$B$_2$ are the most stable new carbon phases theoretically predicted so far. All these four new carbon allotropes are transparent superhard carbon phases with values of bulk modulus and hardness comparable to that of diamond. \\
\section*{Acknowledgments}
This work is supported by the National Natural Science Foundation of China (Grant Nos. 11074211, 10874143 and 10974166), the Cultivation Fund of the Key Scientific and Technical Innovation Project, the Program for New Century Excellent Talents in University (Grant No. NCET-10-0169), and the Scientific Research Fund of Hunan Provincial Education Department (Grant Nos. 10K065, 10A118, 09K033) \\

\end{document}